\documentstyle[12pt]{article}

\newlength{\dinwidth}
\newlength{\dinmargin}
\newcommand{\resection}[1]{\setcounter{equation}{0}\section{#1}}

\begin{document}

\newcommand{\be}{\begin{equation}}
\newcommand{\ee}{\end{equation}}
\newcommand{\bea}{\begin{eqnarray}}
\newcommand{\eea}{\end{eqnarray}}
\newcommand{\nn}{\nonumber}
\newcommand{\dd}{\displaystyle}
\vspace*{1cm}
\begin{center}
  \begin{Large}
  \begin{bf}
NON LEPTONIC TWO BODY DECAYS\\
OF $B$ MESONS$^*$\\
  \end{bf}
  \end{Large}
  \vspace{1cm}
  \begin{large}
{\it A. Deandrea} \\
  \end{large}
{ \it D\'epartement de Physique Th\'eorique, Universit\'e de Gen\`eve}\\
  \vspace{5mm}
\end{center}
\begin{quotation}
\vspace*{3.5cm}
\begin{center}
  \begin{large}
  \begin{bf}
  ABSTRACT
  \end{bf}
  \end{large}
\end{center}
  \vspace{5mm}
\noindent
Within the framework of the factorization approximation, two body non-leptonic
decays of $B$ mesons are related to semileptonic matrix elements and
form factors, evaluated with an effective lagrangian incorporating both chiral
and heavy quark symmetries. Using semileptonic $D$-decay data, estimates for
nonleptonic processes are obtained.
\end{quotation}
  \vspace{2cm}
\begin{center}
UGVA-DPT 1993/10-839\\\
hep-ph/9310334\\\
October 1993
\end{center}
\vspace{2cm}
\noindent
$^*$ Contribution to the proceedings of the Advanced Study Conference on Heavy
Flavours, Pavia, September 3-7. Partially supported by the Swiss National
Science Foundation.

\newpage
\resection{Introduction}
Nonleptonic decays of heavy mesons give us the possibility of exploring
the most interesting and difficult aspects of QCD, in fact nonleptonic
processes are complicated by hard gluon exchanges between the hadronizing
quarks, quark rearrangement and long range effects. The usual assumption is
to use operator product expansion and incorporate long range QCD effects in the
hadronic matrix element of local four quark operators. The calculation makes
use of an effective $\Delta B=1$ hamiltonian \cite{Buras}:
\bea
H_{NL}=\frac{G_F}{\sqrt{2}} \left[ V_{ub}V^*_{uq} (c_1 O_1^u+c_2 O_2^u)
\right. & + &
V_{cb}V^*_{cq} (c_1 O_1^c + c_2 O_2^c)+  \nn \\
 & -& \left. V_{tb}V^*_{tq} \sum^6_{i=3} c_i O_i \right] +h.c.
\label{1}
\eea
where $q=d,s$; $c_i$ are the Wilson coefficients that take into account the
evolution from the scale of the $W$ to the scale $\mu$ of the process under
consideration; $O_1^{u,c}$, $O_2^{u,c}$ are the operators:
\bea
O_1^u=({\bar u}b)_{V-A} ({\bar q}u)_{V-A} \, \, &\,& \,\,
O_1^c=({\bar c}b)_{V-A} ({\bar q}c)_{V-A}  \nn \\
O_2^u=({\bar q}b)_{V-A} ({\bar u}u)_{V-A} \, \, &\,& \,\,
O_2^c=({\bar q}b)_{V-A} ({\bar c}c)_{V-A}
\label{2}
\eea
$O_i$ ($i=3, \ldots 6$) are the so-called penguin operators.
In the previous formulae $q=d,s$ and $({\bar u}b)_{V-A}={\bar
u}\gamma^\mu(1-\gamma_5)b$.
Using factorization and discarding colored current operators
enables one to write the four quark operators in term of more easily
calculable products of current operators, with a coefficient $a_i$ replacing
the Wilson coefficient $c_i$. The link among $a_i$ and $c_i$ is given, for
example for $i=1$, by $a_1=c_1+c_2/N_c$ where the term containing $c_2$ comes
 from the Fierz reordering of $O_2$, with the color octet term arising from
this
transformation discarded. However, as discussed by \cite{BSW}
and \cite{desphande}, the rule of discarding the operators with colored
currents while applying the vacuum saturation is ambiguous and unjustified.
This is a reason, among many others, to make the choice \cite{BSW},
\cite{neubertstech} to treat $a_1$ and the analogous parameter $a_2$,
multiplying $O_2$ as free parameters. Let us recall
that the analysis of $D$ non leptonic decays leads to the empirical finding
$a_1 \approx c_1 ,\; a_2 \approx c_2$ \cite{BSW}.
There has been some recent theoretical effort to understand the empirical rule
of ``discarding the $1/N_c$ term", but these analyses are not conclusive and
apply to particular cases and specific kinematics. So it should not be
surprising that $a_i$ values obtained from a fit turn out to be different from
those expected naively. On the other hand one should always keep in mind that
an ``effective approach'' such as that proposed in the following, that is
factorization with $a_i$ coefficients fitted from data, is only a first
exploratory attempt and numbers obtained in this way should be trusted in most
cases only as order of magnitude estimates.

\resection{Semileptonic form factors}

The idea behind  the factorization approximation is that hadronization
appears only after the amplitude takes the form of a product of matrix
elements of quark currents which are singlets in color, thus allowing
for approximate deductions from semileptonic processes. Different
kinematical situations may suggest that factorization may apply better
to some non leptonic processes rather than to others.
For instance one intuitively expects that it may work better when a color
singlet current directly produces an energetic  meson easily escaping
interaction with the other quarks. Independently of this, various other effects
such as more or less strong role of long-distance contributions, including
final state interactions effects, of small annihilation terms, more or less
sensitivity to choice of the scale, etc., may suggest that the
simultaneous application of the factorization approximation to different
processes must be subject to detailed qualifications. Unfortunately at the
present stage of the subject one is forced to first collect informations
by comparing a very rough procedure to the available data.

According to what already said, in the factorization approximation, two body
non leptonic decays of $B$ and $B_s$ mesons are obtained by the semileptonic
matrix elements of the weak currents between different mesons. A suitable
form is
\be
<P(p')|V^{\mu}|B(p)> =
 \big[ (p+p')^{\mu}+\frac{M_P^2-M_B^2}{q^2} q^{\mu}\big]
F_1(q^2) -\frac {M_P^2-M_B^2}{q^2} q^{\mu} F_0(q^2)
\label{10}
\ee
\bea
<V (\epsilon,p')| &(&V^{\mu}-A^{\mu})|B(p)> =
\frac {2 V(q^2)} {M_B+M_V}
\epsilon^{\mu \nu \alpha \beta}\epsilon^*_{\nu} p_{\alpha} p'_{\beta} \nn\\
&+& i  (M_B+M_V)\left[ \epsilon^*_\mu -\frac{\epsilon^* \cdot q}{q^2}
q_\mu \right] A_1 (q^2) \nn\\
& - & i \frac{\epsilon^* \cdot q}{(M_B+M_V)} \left[ (p+p')_\mu -
\frac{M_B^2-M_V^2}{q^2} q_\mu \right] A_2 (q^2) \nn \\
& + & i \epsilon^* \cdot q \frac{2 M_V}{q^2} q_\mu A_0 (q^2)
\label{11}
\eea
where $P$ is a light pseudoscalar meson and $V$ a light vector meson, $q=p-p'$,
\be
A_0 (0)=\frac {M_V-M_B} {2M_V} A_2(0) + \frac {M_V+M_B}
{2M_V} A_1(0)
\label{12}
\ee
and $F_1 (0)=F_0 (0)$.

For the $q^2$ dependence of all the form factors we have assumed a simple pole
formula $F(q^2)= F(0)/(1-q^2/M_P^2)$ with the pole mass $M_P$ given by the
lowest lying meson with the appropriate quantum numbers ($J^P=0^+$ for
$F_0$, $1^-$ for $F_1$ and $V$, $1^+$ for $A_1$ and $A_2$, $0^-$ for $A_0$).
The values of the form factors at $q^2=0$, are given by the study of the
semileptonic decays as performed in \cite{noi}. For the numerical values of the
form factors and of the pole masses used in this calculation we refer the
reader to \cite{nonlep}.

\resection{Numerical results}

Let us evaluate the coefficient $a_1$. At present, with data nowadays
available, the best way to determine it, is to
consider ${\bar B}^0$ decays into $D^{*+} \pi^-$, $D^{*+} \rho^-$,
$D^{+} \pi^-$, $D^{+} \rho^-$ final states. In order to use the experimental
data we need an input for current matrix elements between $B$ and $D, D^*$
states.

In \cite{noi} we did not consider $B \to D$ and $B \to D^*$ transitions in
the heavy quark effective theory; this subject has been investigated
by several authors and we rely on their work to compute the corresponding non
leptonic decay rates. The relevant matrix elements at leading order in
$1/M_Q$ are \cite{georgi}
\be
<D(v')|V^{\mu}(0)|B(v)> =
\sqrt{M_B M_D} \xi (w) [v+v']^\mu
\label{13}
\ee
\bea
& & <D^* (\epsilon,v')| (V^{\mu}-A^{\mu})|B(v)>  = \nn \\
& &\sqrt{M_B M_D} \xi (w)
\left[ - \epsilon^{\mu\lambda\rho\tau} \epsilon^*_\lambda v_\rho v'_\tau +
i(1+ v \cdot v')\epsilon^{*\mu} -i(\epsilon^* \cdot v)v'^{\mu} \right]
\label{14}
\eea
where $w=v \cdot v'=(M_B^2+M_D^2-q^2)/2 M_B M_D$, $v$ and $v'$ are the
meson velocities, and $\xi (w)$ is the Isgur-Wise form factor. $\xi (w)$ has
been computed by QCD sum rules \cite{sumrul}, \cite{sumrul1}, potential models,
and estimated phenomenologically \cite{neubertstech}.  We shall take
for it the expression \cite{neubert3}
\be
\xi (w) =\left( \frac{2}{1+w} \right) \exp \left[
- (2 \rho^2-1)\frac{w-1}{w+1}\right]
\label{14bis}
\ee
which reproduces rather well the semileptonic data \cite{data} with $\rho
\simeq 1.19, V_{cb}=0.04,\tau_B=1.4 ps$. We stress that we have
chosen to work at the leading order in $1/M_Q$, which is why we have not
introduced the non leading form factors discussed e.g. in \cite{luke},
\cite{neubert}.

{}From the new CLEO data \cite{stone}
$BR({\bar B}^0 \to D^+ \pi^-) = (2.2 \pm 0.5 )\times 10^{-3}$,
$BR({\bar B}^0 \to D^{*+} \pi^-) = (2.7 \pm 0.6 )\times 10^{-3}$ ,
$BR({\bar B}^0 \to D^+ \rho^-) = (6.2  \pm 1.4 )\times 10^{-3}$ and
$BR({\bar B}^0 \to D^{*+} \rho^-) = (7.4  \pm 1.8 )\times 10^{-3}$ one gets
\be
|a_1| \simeq 1.0
\ee
Let us consider now a class of decays that depend only on the parameter $a_2$.
Recent data from CLEO Collaboration \cite{stone} allow for a determination of
this parameter. From $BR(B \to K J/\psi)=(0.10 \pm 0.016) \times 10^{-2}$,
$BR(B \to K^* J/\psi)=(0.19 \pm0.036) \times 10^{-2}$ and
$BR(B \to K \psi(2s))=(0.10 \pm 0.036) \times 10^{-2}$ we obtain
\be
|a_2| \simeq 0.27
\ee
We have now to determine the relative sign between $a_1$ and $a_2$. The new
CLEO data \cite{stone}
$BR(B^- \to D^0 \pi^-) = (4.7 \pm 0.6 )\times 10^{-3}$,
$BR(B^- \to D^{*0} \pi^-) = (5.0 \pm 1.0 )\times 10^{-3}$ ,
$BR(B^- \to D^0 \rho^-) = (10.7 \pm 1.9 )\times 10^{-3}$ and
$BR(B^- \to D^{*0} \rho^-) = (14.1 \pm 3.7 )\times 10^{-3}$,
allow to conclude that the ratio $a_2/a_1$ is positive. Clearly this result
depends on the relative phase of the hadronic matrix elements.
We assume (analogously to \cite{BSW}) that for a
decay $B \to M_1 M_2$ ($M_1$ and $M_2$ scalar or vector mesons) the phase
among the two products of matrix elements in the factorization approximation is
the one determined under the assumption of spin and flavour symmetry in
the meson spectrum. Of course these symmetries are (even
badly) broken in many decays; one should therefore be aware of the
possibility to have a different phase between the two terms in such cases.

We now come to our numerical results. For the values of $f_P$ and $f_V$ used
in computing the rates we refer to the values and the discussion in
\cite{nonlep}.

Let us now consider some example of branching ratios obtained in this
framework. For a more exhaustive list see \cite{nonlep}. Let us stress that the
quoted errors refer only to the experimental input regarding
semileptonic form factors. The theoretical uncertainty is difficult to evaluate
and in general we should regard these numbers only as a first estimate.

\begin{table}[here]
\centering
\caption{{ \it Predicted widths and branching ratios for ${\bar B}^0$ decays.
We use $\tau_{{\bar B}^0}=14 \times 10^{-13} s$, $V_{ub}=0.003$, $V_{cb}=0.04$.
The quoted errors come from the uncertainties on the form factors. Theoretical
errors are not included.}}
\begin{tabular}{l c c }
& & \\
\hline \hline
Process & $\Gamma$ in $10^{12} s^{-1}$& Br \\ \hline
$\pi^+ \pi^-$ & $1.5 a_1^2 |V_{ub} V^*_{ud}|^2$ & $(1.8\pm 0.8)
\times 10^{-5}$ \\ \hline
$\pi^+ \rho^-$ & $4.0 a_1^2 |V_{ub} V^*_{ud}|^2$ & $(4.8\pm 2.9)
\times 10^{-5}$ \\ \hline
$\rho^+ \pi^-$ & $0.3 a_1^2 |V_{ub} V^*_{ud}|^2$ & $(3.6\pm 7.3)
\times 10^{-6}$ \\ \hline
$\rho^+ \rho^-$ & $1.1 a_1^2 |V_{ub} V^*_{ud}|^2$ & $(1.3\pm 2.1)
\times 10^{-5}$ \\
\hline \hline
$\pi^+ {\bar D}_s^-$ & $6.8 a_1^2 |V_{ub} V^*_{cs}|^2$ & $(8.1\pm 3.6)
\times 10^{-5}$ \\ \hline
$\pi^+ {\bar D}_s^{*-}$ & $5.1 a_1^2 |V_{ub} V^*_{cs}|^2$ & $(6.1\pm 2.6)
\times 10^{-5}$ \\ \hline
$\rho^+ {\bar D}_s^-$ & $1.0 a_1^2 |V_{ub} V^*_{cs}|^2$ & $(1.2\pm 2.4)
\times 10^{-5}$ \\ \hline
$\rho^+ {\bar D}_s^{*-}$ & $3.8 a_1^2 |V_{ub} V^*_{cs}|^2$ & $(4.5\pm 2.9)
\times 10^{-5}$ \\ \hline
\hline
$\pi^+ D^-$ & $5.4 a_1^2 |V_{ub} V^*_{cd}|^2$ & $(3.3\pm 1.5)
\times 10^{-6}$ \\ \hline
$\pi^+ D^{*-}$ & $4.0 a_1^2 |V_{ub} V^*_{cd}|^2$ & $(2.5\pm 1.1)
\times 10^{-6}$ \\ \hline
$\rho^+ D^-$ & $0.8 a_1^2 |V_{ub} V^*_{cd}|^2$ & $(4.9\pm 9.8)
\times 10^{-7}$ \\ \hline
$\rho^+ D^{*-}$ & $2.8 a_1^2 |V_{ub} V^*_{cd}|^2$ & $(1.7\pm 1.2)
\times 10^{-6}$ \\ \hline
\hline
$K^0 J/\psi$ & $7.1 a_2^2 |V_{cb} V^*_{cs}|^2$ & $(1.1\pm 0.6)
\times 10^{-3}$ \\ \hline
$K^0 \psi (2s)$ & $2.4 a_2^2 |V_{cb} V^*_{cs}|^2$ & $(0.37\pm 0.19)
\times 10^{-3}$ \\ \hline
$K^{*0} J/\psi$ & $10.4 a_2^2 |V_{cb} V^*_{cs}|^2$ & $(1.6\pm 0.5)
\times 10^{-3}$ \\ \hline
$K^{*0} \psi (2s)$ & $4.8 a_2^2 |V_{cb} V^*_{cs}|^2$ & $(0.74\pm 0.23)
\times 10^{-3}$ \\ \hline
\hline
$\pi^0 D^0$ & $2.6 a_2^2 |V_{cb} V^*_{ud}|^2$ & $(4.1\pm 1.8)
\times 10^{-4}$ \\ \hline
$\pi^0 D^{*0}$ & $2.1 a_2^2 |V_{cb} V^*_{ud}|^2$ & $(3.3\pm 1.4)
\times 10^{-4}$ \\ \hline
$\eta D^0$ & $0.7 a_2^2 |V_{cb} V^*_{ud}|^2$ & $(1.1\pm 0.5)
\times 10^{-4}$ \\ \hline
$\eta D^{*0}$ & $0.5 a_2^2 |V_{cb} V^*_{ud}|^2$ & $(8.6\pm 4.2)
\times 10^{-5}$ \\ \hline
$\rho^0 D^0$ & $0.4 a_2^2 |V_{cb} V^*_{ud}|^2$ & $(6.1\pm 12.2)
\times 10^{-5}$ \\ \hline
$\rho^0 D^{*0}$ & $1.4 a_2^2 |V_{cb} V^*_{ud}|^2$ & $(2.2\pm 1.4)
\times 10^{-4}$ \\ \hline
\hline
${\bar K}^0 D^0$ & $4.4 a_2^2 |V_{cb} V^*_{us}|^2$ & $(3.5\pm 1.8)
\times 10^{-5}$ \\ \hline
${\bar K}^0 D^{*0}$ & $3.4 a_2^2 |V_{cb} V^*_{us}|^2$ & $(2.8\pm 1.3)
\times 10^{-5}$ \\ \hline
${\bar K}^{*0} D^0$ & $0.5 a_2^2 |V_{cb} V^*_{us}|^2$ & $(3.6\pm 7.1)
\times 10^{-6}$ \\ \hline
${\bar K}^{*0} D^{*0}$ & $2.3 a_2^2 |V_{cb} V^*_{us}|^2$ & $(1.9\pm 1.3)
\times 10^{-5}$ \\ \hline
\hline
$\pi^0 J/\psi$ & $4.5 a_2^2 |V_{cb} V^*_{cd}|^2$ & $(3.7\pm 1.6)
\times 10^{-5}$ \\ \hline
$\eta J/\psi$ & $1.2 a_2^2 |V_{cb} V^*_{cd}|^2$ & $(1.0\pm 0.4)
\times 10^{-5}$ \\ \hline
$\rho^0 J/\psi$ & $6.7 a_2^2 |V_{cb} V^*_{cd}|^2$ & $(5.3\pm 1.8)
\times 10^{-5}$ \\ \hline
$\omega J/\psi$ & $6.7 a_2^2 |V_{cb} V^*_{cd}|^2$ & $(5.3\pm 1.8)
\times 10^{-5}$ \\ \hline
\hline
$\pi^0 \pi^0$ & $0.73 a_2^2 |V_{ub} V^*_{ud}|^2$ & $(6.4\pm 3.2)
\times 10^{-7}$ \\ \hline
$\eta \eta$ & $0.07 a_2^2 |V_{ub} V^*_{ud}|^2$ & $(6.1\pm 3.0)
\times 10^{-8}$ \\ \hline
$\pi^0 \rho^0$ & $1.6 a_2^2 |V_{ub} V^*_{ud}|^2$ & $(1.4\pm 0.7)
\times 10^{-6}$ \\ \hline
$\eta \rho^0$ & $0.18 a_2^2 |V_{ub} V^*_{ud}|^2$ & $(1.5\pm 1.4)
\times 10^{-7}$ \\ \hline
$\rho^0 \rho^0$ & $0.53 a_2^2 |V_{ub} V^*_{ud}|^2$ & $(4.5\pm 7.0)
\times 10^{-7}$ \\ \hline
$\pi^0 \omega$ & $0.54 a_2^2 |V_{ub} V^*_{ud}|^2$ & $(4.6\pm 4.3)
\times 10^{-7}$ \\ \hline
\hline
\end{tabular}
\end{table}

\newpage
\noindent
{\bf Acknowledgements }\\

I wish to thank N.~Di Bartolomeo, R.~Gatto and G.~Nardulli, for the pleasant
collaboration regarding the work on which this talk is based; R.~Casalbuoni
and F.~Feruglio for many useful discussions on this subject and finally I
acknowledge a discussion with G.~Martinelli and A.~Falk at this conference.
This work has been supported in part by the Swiss National Foundation.

\end{document}